\documentclass[a4paper,11pt,oneside,final]{article}

\usepackage[utf8]{inputenc}
\usepackage[T1]{fontenc}
\usepackage{amsmath}
\usepackage{amsfonts}
\usepackage{amssymb}
\usepackage{graphicx}
\usepackage{hyperref}
\usepackage[top=2.5cm, bottom=2.5cm, left=2.5cm, right=2.5cm]{geometry}

\title{Photonic jet: key role of injection for etchings with a shaped optical fiber tip}
\author{Robin Pierron\footnotemark[2] \footnotemark[1] , Julien Zelgowski\footnotemark[2] , Pierre Pfeiffer\footnotemark[2] , Joël Fontaine\footnotemark[3] , and Sylvain Lecler\footnotemark[2]}

\date{Article published in \textit{Optics Letters} 42, 14 (2017) \\ 
https://doi.org/10.1364/OL.42.002707}

\begin{document}

\maketitle

\footnotetext[2]{ICube Laboratory, University of Strasbourg, CNRS UMR 7357, 300 Bd. Sébastien Brant, 67412 Illkirch, France}
\footnotetext[3]{ICube Laboratory, INSA Strasbourg, CNRS UMR 7357, 24 Bd. de la Victoire, 67084 Strasbourg, France}
\footnotetext[1]{Corresponding author: robin.pierron@unistra.fr}

\begin{abstract}
We demonstrate the key role of the laser injection into a multimode fiber to obtain a photonic jet (PJ). PJ, a high concentrated propagating beam with a full width at half maximum smaller than the diffraction limit, is here generated with a shaped optical fiber tip using a pulsed laser source (1064~nm, 100~ns, 35~kHz). Three optical injection systems of light are compared. For similar etched marks on silicon with diameters around 1~$\mu$m, we show that the required ablation energy is minimum when the injected light beam is close to the fundamental mode diameter of the fiber. Thus, we confirm experimentally that to obtain a PJ out of an optical fiber, light injection plays a role as important as that of the tip shape, and therefore the role of the fundamental mode in the process.  
\end{abstract}

Direct laser etching at sub--wavelength scale remains a challenge. Numerical simulations have shown that the light can be concentrated in the near field with a full width at half maximum (FWHM) smaller than the diffraction limit \cite{Chen,Lecler} and with a power density significantly higher than the incident wave \cite{Lecler,Heifetz,Abdurrochman}. This high concentrated propagating beam is known as a photonic jet (PJ). It has been initially obtained at the output of dielectric particles: cylinders \cite{Chen,Itagi}, spherical particles \cite{Lecler,Li} or square particles \cite{Pacheco-Pena,Liu}.
Experimental demonstrations have been performed at different pulse widths and wavelengths with microspheres deposited directly on the sample either randomly \cite{Munzer,Abdurrochman} or periodically organized \cite{Wu,Guo,Grojo}. Yet, the spheres are not easy to manipulate and they are polluted after the first irradiation due to their contact with the sample. One solution is to trap the particles by an optical tweezer \cite{Mcleod}; but the technique is not easy to implement in an industrial process. 

In the microwaves frequencies, the PJ emerging from a planar waveguide with a shaped tip has been demonstrated \cite{Ounnas}. In the optical range, spheres put at the end of a core--etched fiber \cite{Aouani} or a hollow--core photonic crystal fiber \cite{Ghenuche} have been used to improve the sensitivity in fluorescence correlation spectroscopy. Yet, the applications are limited to low energies otherwise the fiber end is burned or the spheres are ejected. Another solution, that we have developed, is to use an optical fiber with a shaped tip \cite{Zelgowski,Pierron}. 

In this letter, several laser--fiber injection systems have been experimented. The injected light power necessary to produce ablation on a substrate has been explored. The tests have been performed on silicon using a near-infrared nanosecond pulsed PJ at the end of a multimode shaped fiber tip. The role of the fundamental mode in the photonic jet phenomenon out of the optical fiber has been experimentally confirmed and the key role of injection in the process has been demonstrated.

%\section{Experimental details}

The experimental setup consists in a laser source, a system of injection into the fiber and a 3--axis controlled etching platform. The laser is pulsed and emits at a central wavelength ($\lambda$) of 1064~nm, the pulses are 100~ns wide and the repetition rate is 35 kHz. The output beam has a 1/e$^2$ diameter ($D_{laser}$) of 6~mm and the beam quality factor $M^2$ is 1.3. The injection into the fiber system is achieved using an optical component focusing on a SMA connector whose position is controlled by a XYZ stage. The etching platform consists in an optical fiber on a motorized Z--axis stage and the sample is placed perpendicularly to the shaped fiber tip on a motorized XY stage. The experiments are carried out at ambient atmosphere conditions. 

A  multimode 100/140 step--index silica fiber was used with a numerical aperture (NA) of 0.22, a cladding refractive index ($n_g$)  of 1.457  and a  core refractive index ($n_c$) of 1.440. The sample-tip distance is measured and monitored using a camera with a telecentric objective (5x magnification). The setup is controlled by LabView.

The laser--to--fiber coupling efficiency ($\eta$) is defined as the ratio of the power at the fiber tip end and the power available at the laser output. A high sensitivity power detector (Ophir Vega with the 12A--P sensor) has been used for its determination. 

Optical components with different focal lengths, numerical apertures and aberration corrections have been tested: an achromatic doublet with a focal length ($f$) of 19 mm, a 5x objective microscope with $f$~=~40~mm and an aspheric lens with $f$~=~75~mm. 

A monocrystalline silicon wafer with a passive layer has been used as test sample. The laser punctual ablations have been characterized with an optical microscope with a lateral resolution of 610~nm (Zeiss 50x focusing objective, NA~=~0.55).

\begin{figure}[htbp]
\centering
\includegraphics[width=1.0\linewidth]{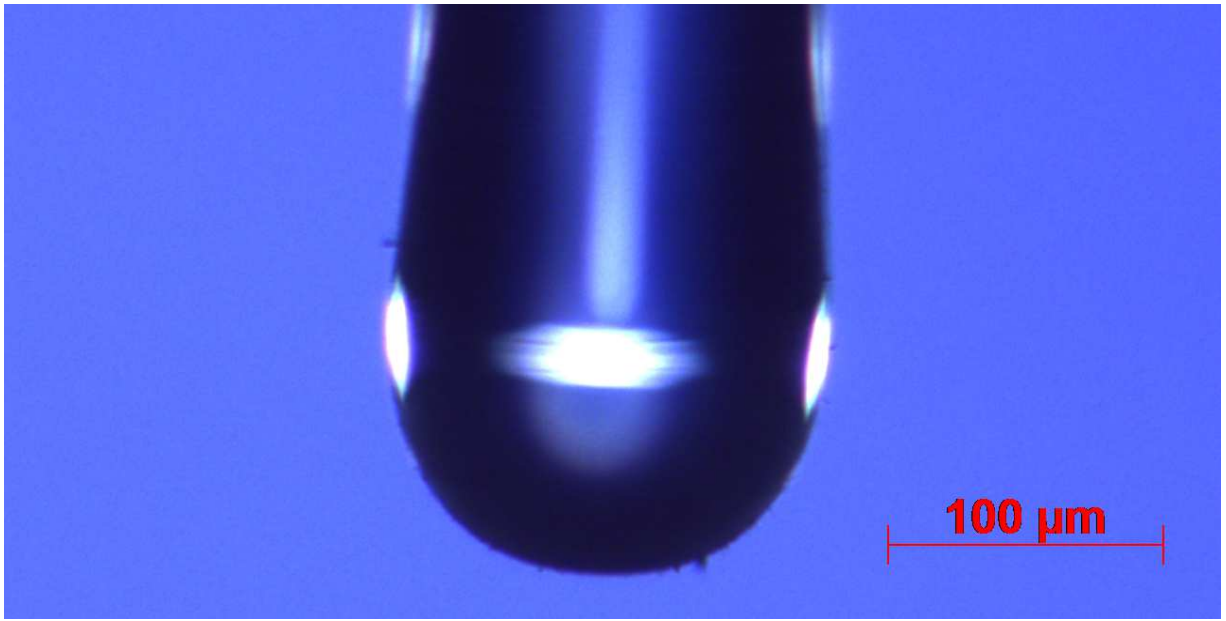}
\caption{Multimode 100/140 optical fiber with a shaped tip. Working distance of 100~$\pm$~2~$\mu$m.}
\label{fig:fiber_tip}
\end{figure}

Numerically, PJs with a FWHM smaller than a half wavelength can be achieved when the maximum PJ intensity is close to the particle \cite{Chen,Lecler} or close to the fiber tip end \cite{Zelgowski}, that is from zero to few wavelengths, which is not practical for an etching process. The tip has been designed to achieve a PJ with a minimum FWHM around 1~$\mu$m at a distance of 100~$\mu$m of the tip in order to avoid any disruption from redeposited melt material. Its shape is described thanks to a Bézier curve set by a base radius ($a$~=~50~$\mu$m), a tip length ($b$ = 63 $\mu$m) and a Bézier weight ($w_0$~=~1) \cite{Zelgowski}. The tip (cf. Fig. \ref{fig:fiber_tip}) has been achieved by the LovaLite company using an electric discharge based thermoforming technique.

%\section{Results and discussion}

\begin{figure}[htbp]
\centering
\includegraphics[width=1.0\linewidth]{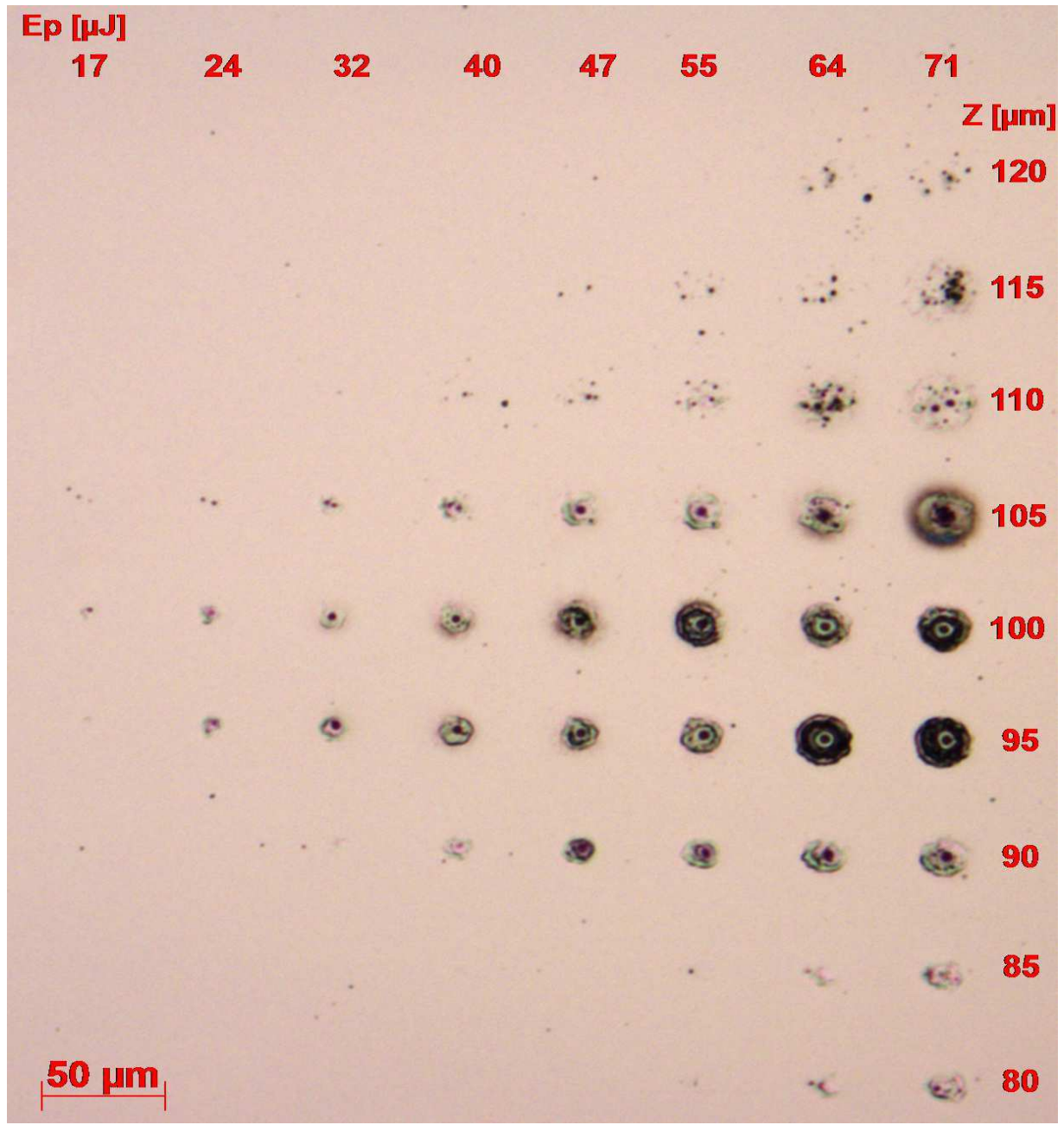}
\caption{Etchings on silicon depending on the sample--tip distance (Z) and the injected power (Ep). Aspheric lens ($f$~=~75~mm). 35 pulses for each PJ etching.}
\label{fig:pt_foc}
\end{figure}

Conventionally, the maximum laser--fiber coupling efficiency can be achieved if the numerical aperture of the focusing optical system (NA$_{foc}$) is lower than the numerical aperture of the multimode fiber (NA). In the paraxial approximation, the NA$_{foc}$ is given by: 
\begin{equation}
\text{NA}_{foc} \simeq \dfrac{D_{laser}}{2f}
\end{equation}
where $D_{laser}$ is the laser beam diameter at 1/e$^2$ and $f$ is the focal length of the optical injection system.
However, in our case, we have considered as optimum the case when the minimum injected power was required to ablate. To find this minimum, the PJ distance has been determined. It corresponds in our etching process to the distance between the tip end and the maximum PJ intensity.  A 2D array of points has been etched (cf. Fig \ref{fig:pt_foc}). Each point has been performed with 35 pulses. On each line the fiber--sample distance was constant, only the power changed; on a column the power was constant, only the distance between the tip and the silicon wafer changed. Different etching sizes appear. The PJ distance can be determined: that is the distance for which the smallest etching is achieved at the smallest power.

As expected, for the three different optical components, a working distance of 100~$\pm$~2~$\mu$m has been found with the smallest etched marks (PJ$_{\varnothing}$) around 1~$\mu$m. However different minimum energies (E$_p$) have been required to ablate (cf. Table \ref{tab:results}). 

\begin{table}[htbp]
\centering
\caption{Minimum pulse energy (E$_p$) injected to begin to etch the silicon for different optical focusing systems: an achromatic doublet ($f$~=~19~mm), a 5x objective microscope ($f$~=~40~mm), an aspheric lens ($f$~=~75~mm). Numerical aperture of focusing system NA$_{foc}$. Spot diameter at the focus point ($D_{foc}$). Laser--fiber coupling efficiency ($\eta$). Etching diameters on silicon (PJ$_{\varnothing}$)}
\begin{tabular}{cccccc}
\hline
$f$ [mm] & NA$_{foc}$ & $D_{foc}$ [$\mu$m] & $\eta$ [\%] & PJ$_{\varnothing}$ [$\mu$m]  & E$_{p}$ [$\mu$J] \\
\hline
19 & 0.16 & 6 & 76  & 1.2 & 36 \\
40 & 0.08 & 12 & 64 & 1.5 & 23 \\
75 & 0.04 & 23 & 61 & 1.3 & 17 \\
\hline
\end{tabular}
  \label{tab:results}
\end{table}

The results (cf. Table \ref{tab:results}) show that the maximum laser--fiber coupling efficiency ($\eta$ = 76\%) is achieved when the NA$_{foc}$ is close to the NA of the multimode fiber. However, in this case, the minimum energy required to etch (E$_{p}$) is not the smallest. The lowest energy required to etch is reached when the coupling efficiency is only 61\%. Moreover, the results reported in Table \ref{tab:results} show that when $D_{foc}$ = 12 $\mu$m the minimum energy required to ablate is higher than when $D_{foc}$ = 23 $\mu$m. It can be explained by taking into account the intensity distribution in the fiber groups of modes. Indeed, with a shaped optical fiber the photonic jet is due to the fundamental mode; the higher--order modes are focused off--axis on larger ring--shape surfaces \cite{Zelgowski}. The higher is NA$_{foc}$, the more energy is injected in the higher groups of modes and the lower is the power in the fundamental mode of the fiber.  

Smaller pulse energies are required to etch if the spot size at the focus point of the converging light matches with the fundamental mode diameter of the fiber. The diameter of the light at the focus point ($D_{foc}$) for a laser beam is given by \cite{Hachfeld}:  

\begin{equation}
D_{foc} = \dfrac{4}{\pi} \ \dfrac{\lambda f}{D_{laser}} \ M^2
\end{equation}

The diameter of the fundamental mode ($2w$) for a step--index fiber has been given by Marcuse \cite{Marcuse} approximating the fundamental Gaussian mode profile by:

\begin{equation}
\dfrac{w}{a} = 0.65 + \dfrac{1.619}{V^{3/2}} + \dfrac{2.879}{V^6}
\end{equation}
where $a$ is the radius of the fiber and $V=a\cdot 2 \pi/\lambda \cdot \text{NA}$ is the normalized frequency. In our case, the diameter of the fundamental mode (2$w$) is 65~$\mu$m. In this way, when the spot diameter at the focus point was close to the diameter of the fundamental mode, the required injected energy to ablate has been reduced by a factor of two. 

In order to evaluate the ratio of the injected energy really involved in the etching process, the theoretical pulse energy required to reach the silicon ablation threshold has been computed. In this case, all the energy was assumed to be concentrated on the etched surface as if only the fundamental mode was involved. Under our operating conditions (1064~nm, 100~ns, 35~kHz), a threshold fluence of around 10~J/cm$^2$ has been considered (a material heating depth $L_H$ of 20~$\mu$m determined from \cite{Meyer} has been used in \cite{Wang} to evaluate it). If we consider the best case ($f$~=~75~mm, PJ$_{\varnothing}$~=~1.3~$\mu$m, E$_{p}$~=~17~$\mu$J), the theoretical minimum required pulse energy is 0.13 $\mu$J; just over 100 times smaller than the measured one. This factor may be in accordance with the estimation of the mode groups number $M = 46$ ($M = V / \sqrt{2}$ \cite{Marcuse2}), taking into account that knowing the exact energy distribution on the guided modes is difficult. Furthermore, in our process the optical fiber is bent due to its length ($\simeq$ 1.2~m), which introduces mode coupling. 

\begin{figure}[htbp]
\centering
\includegraphics[width=1.0\linewidth]{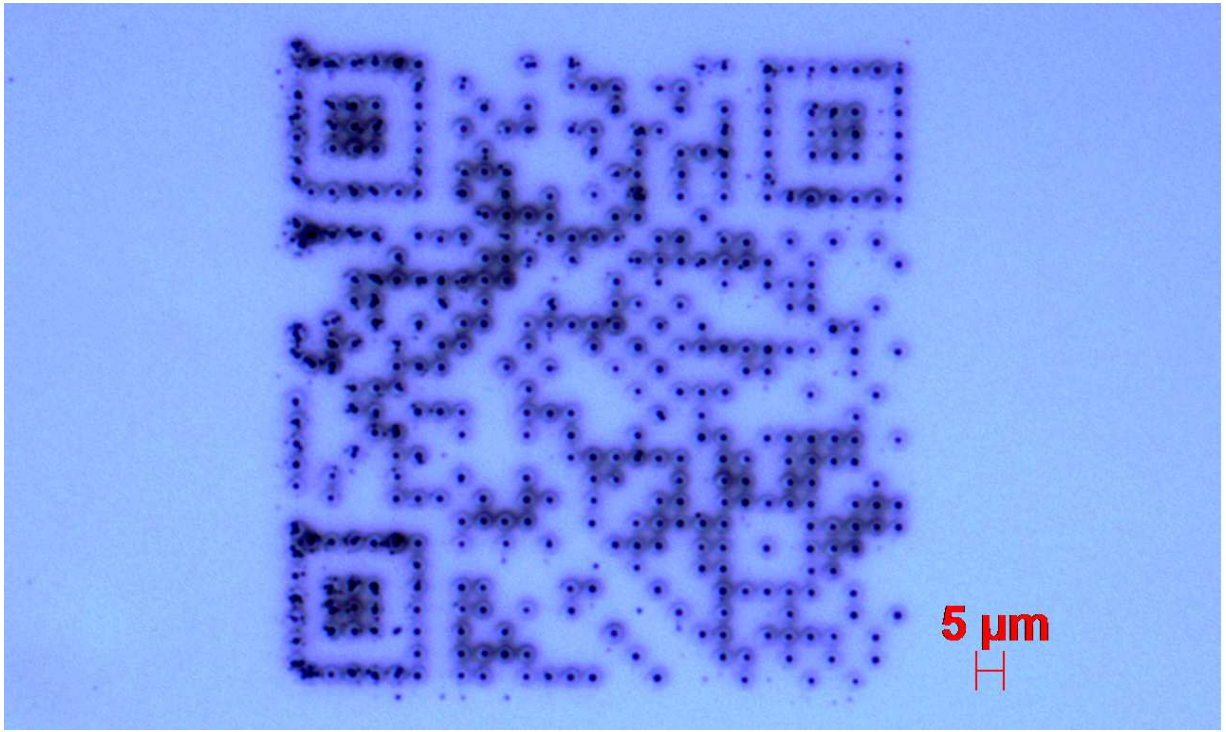}
\caption{QR code matrix (110x110~$\mu$m) on silicon. Each etched mark has a diameter around 1.3~$\pm$~0.6~$\mu$m. Pulse energy of 17~$\mu$J. Aspheric lens ($f$~=~75~mm). 35 pulses for each PJ etching.}
\label{fig:QRcode}
\end{figure}

The laser process is repeatable. The fiber tip is not disturbed by the recast material and can be easily manipulated. The required energy can be minimized by controlling the injection. For example, a QR code has been achieved with a set of micro dots with diameters around 1 $\mu$m etched using only 17 $\mu$J per pulse (cf. Fig. \ref{fig:QRcode}). In the figure, a shading can be observed around some ablated dots. Closer the etching dots, larger the shading around. The dots are spatially closed (5 $\mu$m) and achieved in short time (travel speed of the motorized XY stages: 1 mm/s). It is a heat accumulation effect, due to the high-order modes of the fiber which are spread off-axis \cite{Zelgowski} and do not contribute to the etching. This confirm the importance of the energy distribution on the modes inside the fiber.

%\section{Conclusion}

To conclude, three optical components for the injection of the laser into the fiber have been used with different numerical apertures and spot diameters. The tests have been performed with a 100/140~$\mu$m thermoformed fiber tip. At the lowest numerical aperture (NA$_{foc}$~=~0.04), when the spot diameter was close to the fundamental mode diameter of the fiber, the required energy to ablate has been reduced by a factor of two. Therefore, the experiments corroborate that the photonic jet is mainly due to the fundamental mode \cite{Zelgowski}. This also demonstrates that the required energy can be minimized by controlling the injection to achieve etched marks having diameter smaller than the wavelength with a shaped multimode fiber tip. 

\paragraph{Funding}

SATT Conectus Alsace; LaserJet Project.

\paragraph{Acknowledgment}

The authors would like to thank Gary Argaud for his technical help.

% Bibliography
\bibliography{mybibfile}

%\section*{References}

\bibliographystyle{ieeetr} 

\end{document}